\documentclass[5p,times]{elsarticle}
\usepackage[misc]{ifsym}
\usepackage{amssymb}
\usepackage{amsmath}
\usepackage{multicol}
\usepackage{graphics}
\usepackage{mathtools,cuted}
\usepackage{lipsum}
\usepackage{multirow}
\usepackage{threeparttable}
\usepackage{color}
\usepackage{algorithm}  
\usepackage{algorithmicx}  
\usepackage{algpseudocode}  
\usepackage[colorlinks=true,linkcolor=blue]{hyperref}%
\usepackage[figuresright]{rotating}
\begin{document}

\begin{frontmatter}
 \title{Evolution Features and Behavior Characters of Friendship Networks on Campus Life}
%% \tnotetext[label1]{}
 \author[add1,add2]{Zongkai Yang}
 \author[add1]{Zhu Su\corref{cor1}}
 \author[add1,add2]{Sannyuya Liu\corref{cor1}}
 \author[add1]{Zhi Liu}
 \author[add1]{Wenxiang Ke}
 \author[add1]{Liang Zhao}

 \address[add1]{Natoinal Engineering Laboratory for Educational Big Data, Central China Normal University, Wuhan 430079, China}
 \address[add2]{National Engineering Research Center for E-Learning, Central China Normal University, Wuhan 430079, China}
 \address{[zkyang,suz,lsy5918,zhiliu,liang.zhao]@mail.ccnu.edu.cn; kewenxiang@mails.ccnu.edu.cn}
\cortext[cor1]{Corresponding authors. \\Email addresses: suz@mail.ccnu.edu.cn (Z. Su), lsy5918@mail.ccnu.edu.cn (S. Liu).}

\begin{abstract}
Analyzing and mining students' behaviors and interactions from big data is an essential part of education data mining. Based on the data of campus smart cards, which include not only static demographic information but also dynamic behavioral data from more than $30000$ anonymous students, in this paper, the evolution features of friendship and the relations between behavior characters and student interactions are investigated. On the one hand, four different evolving friendship networks are constructed by means of the friend ties proposed in this paper, which are extracted from monthly consumption records. In addition, the features of the giant connected components (GCCs) of friendship networks are analyzed via social network analysis (SNA) and percolation theory. On the other hand, two high-level behavior characters, orderliness and diligence, are adopted to analyze their associations with student interactions. Our experiment/empirical results indicate that the sizes of friendship networks have declined with time growth and both the small-world effect and power-law degree distribution are found in friendship networks. Second, the results of the assortativity coefficient of both orderliness and diligence verify that there are strong peer effects among students. Finally, the percolation analysis of orderliness on friendship networks shows that a phase transition exists, which is enlightening in that swarm intelligence can be realized by intervening the key students near the transition point.
\end{abstract}

\end{frontmatter}
\section{Introduction}
Social computing has become a promising research area and has attracted much attention. Investigating student behaviors and student interactions at a large scale has always been a huge challenge for traditional educational researchers due to their complexity and uncertainty. For a traditional research paradigm, small-scale follow-up surveys and laboratory tests are the most common methods, generally conducted in the form of questionnaires \citep{Robins2007Handbook,Thea2008, JUNCO2013626}. Nevertheless, the disadvantage of these methods is that the data are too subjective or too limited to obtain reliable results. Thanks to the deep integration of information technology in education, the behavioral data of most students on campus can be collected by mobile phones \citep{Wang2014StudentLife}, online courses \citep{Reich130, Reich34}, WiFi \citep{Zhou2016EDUM}, etc., which provide the potential of large-scale and long-term empirical analyses for researchers. Thus, mining and analyzing the hidden features from these data are extraordinarily important for understanding student behavior patterns and for interpreting a large number of complicated phenomena among learning communities. For example, by analyzing the data in massive open online courses (MOOCs), Brinton et al. \cite{Brinton2015MOOC} found that watching more videos and making more than one pause are two strong indicators for students obtaining excellent academic performance.

The use of scientific methods to quantify student behaviors and describe student interactions is a significant step toward personalized education, which not only helps education administrators quantitatively understand the major factors of excellent/poor performance but also helps students discover the gap with others and obtain a clear understanding of their situations under a macro learning background, thus stimulating their learning interests, enhancing learning effects and improving comprehensive quality. To this end, studies have proposed many methods to analyze and mine student behaviors and student interactions to understand the essential mechanism of macroscopic phenomena and provide early warnings of collective emergencies. For example, social network analysis (SNA), as a powerful tool, has been applied in the educational field due to the ease of describing abundant interaction processes \citep{DEMARCOS2016312}. \cite{Yi2018} proposed orderliness and diligence to quantify student behaviors and demonstrated that these two characters could predict student academic performance.

Most previous studies have focused on analyzing the features of static network topology, but social ties among students are not static and change over time; thus, it is necessary to study the evolution features of friendship networks, which could help us understand how friendships form and disappear. Moreover, student behaviors are always influenced by peers, but the specific influences peers have are still unclear. To solve the above problems, we propose an approach of inferring friend ties to construct evolving friendship networks from personal consumption data on campus and to adopt orderliness and diligence as two important behavior characters to investigate their associations with student interactions.

To construct friendship networks, we propose a theoretical method to infer friend ties from more than $30,000$ student consumption data in university canteens. The reliability of the inferred friendship network is confirmed by comparison with that of self-report friendship data from $42$ students. Evolution features of friendship are investigated by SNA, and the relations between behavior characters and student interactions are investigated by assortativity analysis and percolation analysis. Our main findings include the following: (i) The size of the friendship network declines over time. (ii) The small-world effect and power-law distribution are revealed in friendship networks. (iii) The orderliness and diligence are positively related to student academic performances, and the peer effect and a phase transition of behavior characters are uncovered in friendship networks.

Our main contributions can be summarized as follows:

1. Considering the features of student behavior data, we propose a theoretical framework to determine the critical value of the co-occurrence frequency in various time windows for inferring friend ties. This method is simple, reasonable, and highly accurate.

2. The evolution features of friendship are investigated by analyzing the topological characteristics of four friendship networks and their giant connected components (GCCs) at different times. We have found that students make many friends when they are in a new environment; as time goes on, only like-minded friends remain.

3. The distribution characteristics of orderliness and diligence are explored, as well as their relations to academic performance. Based on friendship networks, the relations between the behavior character and student interactions are investigated by assortativity analysis and percolation analysis, respectively.

The remainder of the article is organized as follows. First, we provide an overview of the literature on SNA and percolation theory. Then, we introduce the materials and methods, followed by the results. The paper ends with a conclusion, limitations and suggestions for further research.

\section{Related studies}
\subsection{Social network analysis}
The interactive relations among people can be described by a network or graph consisting of nodes and links, where the nodes stand for individual actors and links represent relationships among individuals. SNA is able to show, explore and explain the structure character of networks \citep{Empygiri2014Using}, which can help us obtain an in-depth understanding of social phenomena. In fact, SNA can supplement quantitative data analysis to generate the summation of learning results by adding explanations for group dynamics between the subjects \citep{LEE201635}, thus providing the theoretical basis for empirical results.

Investigating the features of a social network among students can provide references for the application of group learning; hence, an increasing number of researchers have applied SNA in exploring the relationship between social ties and academic performance \citep{Dawn2016, Morelli9843}. The results indicated that there is a strong correlation between friends and academic performance \citep{TOPIRCEANU2017171}. For example, \cite{Stadtfeld792} studied how social relationships formed by students who do not know each other and explained their academic success by tracking $226$ undergraduates from the beginning to the end of the academic year. These researchers uncovered that friends can evolve into learning relationships, which demonstrates that the social network is a key factor of academic success. Based on the data of high school students, \cite{Jennifer2012} modeled dynamic networks by means of the Markov model with random behavior individuals and studied the coevolution of the network and behavior. These authors observed that high-achieving students were more likely to become friends with high-achieving students, indicating that academic performance could be improved by changing friendship relationships. \cite{Veenstra2018} used SNA to study the influence of friend relationships on adolescent behavior and demonstrated that friendship plays an important role in shaping adolescent academic achievement and risky behavior. \cite{Kassarnig2018Academic} studied the social network of $538$ undergraduates from smart phone data and discovered that network indicators could better reflect the academic performance of students than individual characteristics and that the network has a strong peer effect.

Despite the advantages of SNA for analyzing and computing the network structure \citep{Buchanan2015A}, it is not always easy to capture any given system as a network since not all systems have an obvious network structure where the interconnections can be obtained from direct observation. Moreover, the collected data may not capture the associations among observed objects leading to a hidden relational structure. Because of the above situations, some researchers have proposed several methods to infer social ties from various data sets \citep{Crandall22436, Tang2011, Sapiezynski2017}. For example, \cite{Eagle15274} inferred friend ties through location and proximity data from mobile phones, and the results demonstrated that it is possible to accurately infer $95\%$ of friend ties based on observational data alone. \cite{Tang2012} proposed a framework for inferring social ties by incorporating social theories into a machine learning model, and an F1-score of $90\%$ was obtained. In addition, the statistical validation method has also been used for the inferring of social ties \citep{Li2014Statistically, Li2014, Li_2014}. For instance, \cite{Liu2017} developed a statistical validation and measured the similarity or relationship among students based on their spatio-temporal co-occurrences, and they found the friendship network is highly assortative by students' attributes such as gender, grade, school and age.

\subsection{Percolation theory}
Percolation theory \citep{Achlioptas1453, Bohman1438} is a theory of random graphs to study the emergence of large-scale connected components of networks on the gradual addition of links/nodes with a connect/active threshold $p$, which is also called bond/site percolation. More specifically, taking site percolation as an example, given a network, we hypothesize that nodes are active with probability $p$. For $p = 0$, nodes are inactive in the network, leading to a disconnected configuration. For $p = 1$, all nodes are active, and the whole clusters in the network are presented. As $p$ varies, the network undergoes a structural transition between these two extreme configurations. Generally, random percolation processes give rise to continuous phase transitions \citep{Dorogovtsev2008}. This finding means that the size of the largest cluster in the network, used as a proxy for the connectivity of the system, increases from the nonpercolating to percolating phases in a smooth fashion \citep{Radicchi2015}.

Percolation theory was proposed by Boardbent and Hammersley in 1956 \citep{Broadbent1957Percolation}. Initially, it was used to describe the random expansion and flow of fluid in random porous media. Since percolation theory is of great practical significance, it is widely applied to explain many physical, chemical and biological phenomena \citep{Goffri2006, BAUHOFER20091486}.

Percolation is easy to formulate for exploring the critical phenomena and rules of group behavior when a percolation transition occurs; thus, percolation theory has also been gradually applied to the social sciences \citep{GOLDSTONE2005424}. For instance, \cite{SOLOMON2000239} investigated the percolation phenomenon of the social network of customers in the media industry. These investigators observed self-organized criticality toward the usual percolation threshold and related scaling behavior by computer simulation on square lattices. \cite{ZHUKOV2018297} applied percolation theory to explore how society formed from individuals to connected groups. \cite{JIANG201810} analyzed the relocation patterns of the manufacturing industry on the network of the Yangtze River Economic Belt using percolation theory and found that percolation transitions exist during the process of industry relocation. \cite{Li669} investigated the percolation transition of traffic networks from real-time traffic data. The results indicated that local congested bottlenecks can lead to a global traffic breakdown; therefore, developing the traffic capacity on these bottlenecks can significantly improve global traffic.

\section{Materials and methods}
\subsection{Data description}
The results presented in this paper are based on data collected from the student card system (SCS) at Central China Normal University (CCNU), Hubei, P. R. China. The system includes three components: the radio-frequency identification (RFID) tag system, campus smart card and database. RFID tags are armed in several locations, such as student canteens, dormitories, libraries, classrooms and stores across the university campuses. Most of the student campus behaviors, such as having meals in canteens, shopping in stores and entering the library, are recorded via a campus smart card check-in and are uploaded to the database.

The work described here is a part of the data we extracted from the SCS database, including behaviors occurring in student canteens and libraries. When students go to the student canteens for dinner, they should have their card scanned to pay for the meal, and the RFID tag system records the following information: \textit{student ID}, \textit{location}, and \textit{timestamp}. Analogously, the library check-in data will be recorded when a student enters the library using his/her card. 

In this paper, we collected approximately $5,602,261$ records of entering the library from September, $2015$ to July, $2019$ and $8,107,001$ records of canteen consumption data from March, $2019$ to July, $2019$. In the data preprocessing, we filtered the invalid data so that the number of student consumption records was less than $10$ times per month. According to our statistical results, the number of valid students is $31,980$ in March, $31,344$ in April, $31,154$ in May, and $29,361$ in June. Moreover, there are $8$ canteens and $167$ windows around the university campus. After collecting these data, we also acquired the grade point average (GPA) for  $28,926$ students from the academic database. Specifically, we calculated $66,874$ diligence values from the records of entering the library, and $40,924$ orderliness values from the records of canteen consumption data, respectively. In order to investigate the relationship between the diligence (or orderliness) and GPA, we obtained $26,753$ (or $17,996$) samples that each student has both the diligence (or orderliness) and GPA. To investigate the periodicity features of friendship networks, we collected the other two semester records of canteen consumption data, and the details are listed in Table \ref{table0}.

\begin{table}
	\caption{ The information of canteen consumption data.}
	\label{table0}
	\centering
	\begin{tabular}{cccc}
		\hline\hline
		&Records&Valid Students\\
		\hline 
		03/18&2171314&28873\\
		04/18&1822634&28089\\
		05/18&2073499&28616\\
		06/18&1911806&27486\\
		\hline
		09/18&2504612&33008\\
		10/18&2201920&32359\\
		11/18&2246927&33293\\
		12/18&2169025&32478\\
		\hline
		03/19&2245626&31980\\
		04/19&2082048&31344\\
		05/19&2038350&31154\\
		06/19&1740977&29361\\
		\hline\hline
	\end{tabular}
\end{table}

In our study, privacy protection was taken quite seriously, and all the students' information was anonymous. Both the student name and student number in our raw data are already pseudonymous. The institutional review board (IRB) from Central China Normal University approved the study.

\subsection{The method of inferring friend ties}{\label{method of constructing network}}
The idea of inferring friend ties is based on the fact that friends often have meals together, and the chances of friends are larger than that of strangers appearing at the same canteen window simultaneously. Therefore, the friendship network can be constructed using the co-occurrence frequency within a period of time. Indeed, it is very important to determine the critical value of the co-occurrence frequency because a smaller frequency might mistake strangers for friends, whereas a larger frequency would omit some real friends. Toward this end, in this section, we derive a probability formula as well as mathematical expectations in theory and calculate certain specific values.

First, we hypothesize that there are $m$ students and $n$ canteen windows in the university, and the co-occurrence chance of any two students is $p$. Then, the probability $P$ and mathematical expectations $E$ that two students co-occur at least $a$ times within $b$ meals can be expressed as:
\begin{equation}\label{eq1}
\begin{split}
P(x \ge a) &= \sum_{x=a}^{b}C_{b}^{x}p^{x}(1-p)^{b-x}\\
E(x \ge a) &= \sum_{x=a}^{b}C_{m}^{2}C_{b}^{x}p^{x}(1-p)^{b-x}
\end{split}
\end{equation} 
In fact, the co-occurrence must satisfy the following two conditions: The first one is that people appear at the same canteen window, and the second one is that they appear in the same time interval. Ignoring personal preferences, people choosing the canteen window can be regarded as a random behavior; thus, the probability of any two students at the same canteen window can be expressed as:
\begin{equation}
 p_{1}=\sum_{i=1}^{n}(\frac{1}{n})^2=\frac{1}{n}. 
\end{equation}

Meanwhile, since most students have meals at a regular time, the time of students appearing in the canteen window follows a normal distribution $N(\mu,\sigma^{2})$. For example, almost all the students had lunch between 11:00 a.m. and 1:00 p.m., especially at 12:00 p.m. Therefore, the probability that two students have meals in the same time interval can be expressed as:
\begin{equation}
 p_2=\sum_{i=1}^{N-1}(\int_{\Delta t_{i}}^{\Delta t_{i+1}}\frac{1}{\sqrt{2\pi}\sigma}e^{-\frac{(x-\mu)^2}{2\sigma^2}}dx)^2. 
\end{equation}

Combining these two conditions, the co-occurrence probability can be written as: $p=p_1\times p_2$. Substituting these relations into \eqref{eq1} gives:

\begin{equation}\label{eq2}
\begin{split}
P(x \ge a)&= C_{b}^{x}\times \\
&\left\lbrace  \frac{1}{n} \sum_{i=1}^{N-1}\left( \int_{\Delta t_{i}}^{\Delta t_{i+1}}\frac{1}{\sqrt{2 \pi}\sigma}e^{-\frac{(y-\mu)^2}{2\sigma^2}}dy\right) \right\rbrace ^{x}\times \\
& \left\lbrace  1-\frac{1}{n} \sum_{i=1}^{N-1}\left( \int_{\Delta t_{i}}^{\Delta t_{i+1}}\frac{1}{\sqrt{2 \pi}\sigma}e^{-\frac{(y-\mu)^2}{2\sigma^2}}dy\right) \right\rbrace ^{b-x}  \\
E(x\ge a)&=C_{m}^{2}C_{b}^{x} \times\\ 
&\left\lbrace  \frac{1}{n} \sum_{i=1}^{N-1}\left( \int_{\Delta t_{i}}^{\Delta t_{i+1}}\frac{1}{\sqrt{2\pi}\sigma}e^{-\frac{(y-\mu)^2}{2\sigma^2}}dy\right) \right\rbrace ^{x}  \times \\
&\left\lbrace  1-\frac{1}{n} \sum_{i=1}^{N-1}\left( \int_{\Delta t_{i}}^{\Delta t_{i+1}}\frac{1}{\sqrt{2 \pi}\sigma}e^{-\frac{(y-\mu)^2}{2\sigma^2}}dy\right) \right\rbrace ^{b-x} 
\end{split}
\end{equation}

\begin{algorithm}
 	\caption{Infer friend ties}
 	\begin{algorithmic}[1]
 		\Require $data1$: a student's canteen consumption data sequence;\\ $data2$: other students' canteen consumption data sequence
 		\Ensure $friendlist$: friend list
 		\State Initialize value	$loc$, $t$, $i \gets 0$, initialize sequence $friend$
 		\State $a$, $b$, $c \gets $ $data1['num','t','loc']$
 		\For{$loc \gets c[0]$ to $c[-1]$}
 		\State $t \gets b[i]$
 		\For {$temp \gets data2[0]$ to $data2[-1]$}
 		\State $j \gets 0$
 		\For {$x$ $\gets$ $temp['loc'][0]$ to $temp ['loc'][-1]$}
 		\State $x\_t \gets temp['t'][j]$
 		\If {$(x == loc)$ \&\& $((x\_t >= t -120)$ \&\& $(x\_t <= t+120))$}
 		\State $ friend.append(temp['num'])$
 		\EndIf
 		\State $j \gets j+1$
 		\EndFor
 		\EndFor
 		\State $i \gets i+1$
 		\EndFor
 		\For{$m \gets friend[0]$ to $friend[-1]$}
 		\If{$len(m) > 5$}
 		\State $friendlist \gets m$
 		\EndIf
 		\EndFor
 		\State\Return{$friendlist$}
 	\end{algorithmic}
 \end{algorithm}

According to the statistical results of our collected data, there are approximately $30,000$ students and $160$ canteen windows in the university. A student would have $90$ meals for one month, and the time of a meal is primarily distributed within $60$ minutes. Here, we assume that students at the same canteen window do co-occur if their consumption time is within two minutes. Thus, the above parameters can be set up as $m=30,000$, $n=160$, $b=90$, $N=60$, $\Delta t = 2$, and $\sigma = 20$ (meaning that $99.74\%$ students have meals within two hours). 

\begin{figure}[ht]
	\centering
	\includegraphics[scale=0.5]{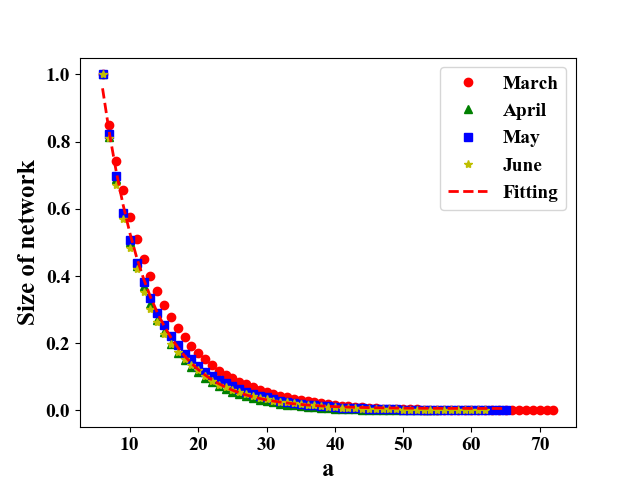}
	\caption{\label{Figure1} (Color online) The network size depends on the co-occurrence frequency $a$ in various months. }
\end{figure}

Using Mathematica software, we calculate some values in the case that $a$ is arranged from $1$ to $9$, as shown in Table \ref{table1}. It is very easy to find that the critical value of the co-occurrence frequency $a_c$ is equal to $5$ because in this case, we cannot find any pair of strangers ($E(x>5) =3.32E-3$); that is, they are massively more likely to be friends if they co-occur at least $5$ times. Therefore, we use this criterion to infer friend ties from student consumption data and construct corresponding friendship networks. The particular algorithm of inferring friend ties is presented in Algorithm $1$. Based on this algorithm, we construct friendship networks in various co-occurrence frequencies $a$. As shown in Fig. \ref{Figure1}, a remarkable exponential function relation is found between the size of the network and the co-occurrence frequency, which approximately satisfies $f(a) \sim 1.89exp(-0.22a+0.076)$.

\begin{table*}
	\centering  
	\caption{The probability $P$ and mathematical expectations $E$ in different encounter frequencies $a$.} 
	\label{table1}
	\begin{tabular}{cccccccccc}
		\hline  
		$a$&1&2&3&4&\textbf{5}&6&7&8&9\\
		\hline 
		$P(x\ge a)$&1.57e-2&1.23e-4&5.74e-7&2.43e-9&\textbf{7.38e-12}&1.84e-14&3.89e-17&5.71e-19&81.14e-22\\
		$E(x \ge a)$&7.08e+6&5.54e+4&2.86e+2&1.10&\textbf{3.32e-3}&8.29e-6&1.75e-8&2.32e-10&5.14e-14\\
		\hline 
	\end{tabular}
	\label{table:stu}  
\end{table*}

\subsection{Orderliness}{\label{orderliness}}
Orderliness is described as the regularity of student behaviors. For example, if the starting times of having lunch for student $A$ always fall into the fixed time range $[12:00, 12:30]$, whereas student $B$ has lunch at any time randomly, then we could say student $A$ has a higher orderliness than student $B$. According to the definition in reference \cite{Yi2018}, the orderliness $O_{\xi}$ can be quantified by an actual entropy $S_{\xi}$, obtained through calculating a time series of a specific behavior, and their relation can be written as $O_{\xi}=-S_{\xi}$. The smaller the actual entropy is, the higher the orderliness is. The actual entropy can be calculated in detail by the following method.

First, the preprocessing of the time series should be conducted. The sequence of our original data is $e = \{ t_1-d_1-L_1,t_2-d_2-L_2,\cdots,t_n-d_n-L_n\}$, where $t_n$ denotes the precise time between $00:01$ and $24:00$, $d_n$ is the exact date, and $L_n$ is one of the canteen windows. Since the precise time within a day is useful for calculating the actual entropy, we keep $t_n$ and finally obtain the new sequence $e^{\prime} = \left\lbrace t_1, t_2, t_3, \cdots, t_n \right\rbrace$. Second, one day is divided into $48$ time slices, each of which spans $30$ minutes and is encoded from $1$ to $48$. Therefore, we obtain a new discrete sequence. For example, $\left\lbrace10:00, 11:00, 12:00, 13:00, 14:00\right\rbrace$ corresponds to the discrete sequence of $\left\lbrace 20, 22, 24, 26, 28\right\rbrace $.

Then, we use the new discrete sequence to calculate the actual entropy, which is defined as:
\begin{equation}
\begin{split}
S_{\xi} = \left( \frac{1}{n}\sum_{i=1}^{n}\Lambda_{i}\right) ^{-1}\ln n
\end{split}
\end{equation} 
where $\Lambda_{i}$ represents the length of the shortest subsequence starting from $t_{i}^{\prime}$ of $\xi$, which never occurred previously. According to the results of reference \cite{XU2019345}, $\Lambda_{i} = n-i+2$ if we could not find the shortest subsequence. For example, for a discrete data sequence $\{16, 23, 35, 16, 23, 35, 33\}$, where $\Lambda_{1} = 1$, $\Lambda_{2}  = 1$, $\Lambda_{3}  = 1$, $\Lambda_{4}  = 4$, $\Lambda_{5}  = 3$, $\Lambda_{6}  =2$, $\Lambda_{7}  = 1$, and $n = 7$, the value of $S_{\xi}$ is $1.048$.

\subsection{Assortativity coefficient}{\label{assortativity}}
The assortativity coefficient $r$ is a Pearson correlation coefficient based on node degrees, which is used to measure the relationship of the connection of node pairs, and its value is between $-1$ and $+1$. A network is assortative if the value of $r$ is positive, where the high-degree nodes tend to connect with high-degree nodes. Instead, when the value of $r$ is negative, high-degree nodes are inclined to connect with low-degree nodes, and the network is disassortative. In addition, the network is neutral and has no degree correlation in the case of $r=0$.

In fact, not only the node degrees, but also other node attributes can be generalized by the assortativity coefficient, which can be calculated by the following equation:
\begin{equation}\label{eq4}
\begin{split}
r = \frac{\sum_{i,j}\left( a_{ij}-\frac{k_{i}k_{j}}{2M}\right) x_{i}x_{j}}{\sum_{i,j}\left( k_{i}\delta_{ij}-\frac{k_{i}k_{j}}{2M}\right) x_{i}x_{j}}
\end{split}
\end{equation} 
Here, $x_i$ (or $x_j$) is the attribute value of node $i$ (or node $j$), $k_i$ (or $k_j$) is the degree of node $i$ (or node $j$), $M$ is the number of edges in the network, $a_{ij}$ is an element of adjacent matrix $A$, in which element $a_{ij}=1$ if nodes $i$ and $j$ are connected and $a_{ij} =0$ otherwise, and $\delta_{ij}$ is the Kronecker delta function, where $\delta_{ij}=1$ if $i=j$, otherwise $\delta_{ij}  = 0$.

\section{Results and discussion}
In this section, we investigate the evolution features and relations between behavior characters and student interactions of friendship networks. To investigate the evolution features of friendship in a semester, four friendship networks are constructed with a 30-day time window, and their topological properties are analyzed. Alternatively, to quantify student behaviors, we adopt two high-level characters, orderliness and diligence, and analyze their distributions and correlations with their GPA to confirm the conclusions of the study \cite{Yi2018}. Next, the relations between behavior characters and student interactions are investigated using the assortativity coefficient and percolation theory, respectively. 

\subsection{Evolution features of friendship}
We design Algorithm 1 to infer friend ties according to the data set character. The algorithm has three nested for loops so that its time complexity is $O(n^3)$ and space complexity is $O(n)$. The experimental computer CPU is Intel i7-9700K, with 32GB of RAM. It runs on the Linux distribution Ubuntu18.04, and the operating environment is Pycharm, Anacoda3 and Python3.7. Due to the limitation of computational ability and RAM, it can support the analysis of up to about $80,000,000$ data records in an academic year to build a friendship network. In order to improve the computational speed and efficiency, we analyze each group of student data (about $6,000,000$ data records) separately on a monthly basis. The calculation time of the friendship network on the experimental computer is $50$ hours.

\begin{figure}[ht]
	\centering
	\includegraphics[scale=0.35]{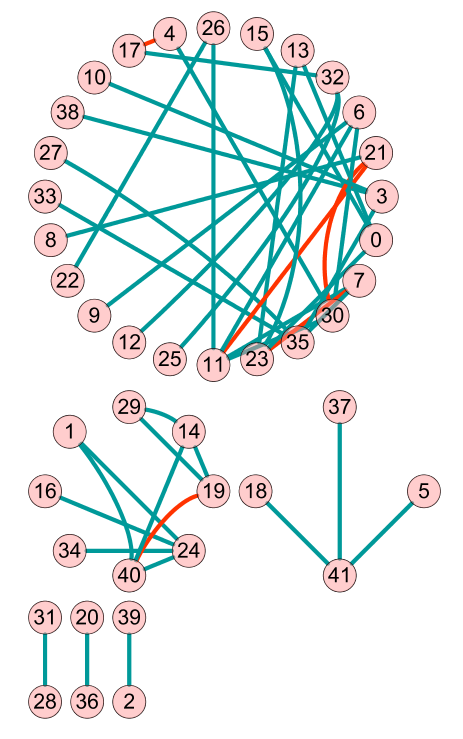}
	\caption{\label{Figure2} (Color online) The inferred and real friend ties among students. Green edges represent the coexistence of these different friend ties, and red edges represent the real friend ties. }
\end{figure}

Using the method in Algorithm 1, we inferred friend ties from monthly consumption records in the university canteens, and constructed four friendship networks from March to June 2019. To verify the reliability of inferring friend ties, we collect self-report friendship data from $42$ student volunteers, who were asked about the information of their friends. According to the results of the self-report, there are $43$ real friend ties among these volunteers. To verify the validity of our inferred results, we selected edges among these volunteers from friendship networks and compared them with real friend ties. As shown in Fig. \ref{Figure2}, only $5$ edges (red lines) cannot be found. The result shows that the hit rate of the friend ties inferred by our method possibly reach $88.4\%$. Therefore, it is reliable that the formations of most edges in the inferred friendship networks are mainly driven by real friend ties.

\begin{figure}[ht]
	\centering
	\includegraphics[scale=0.45]{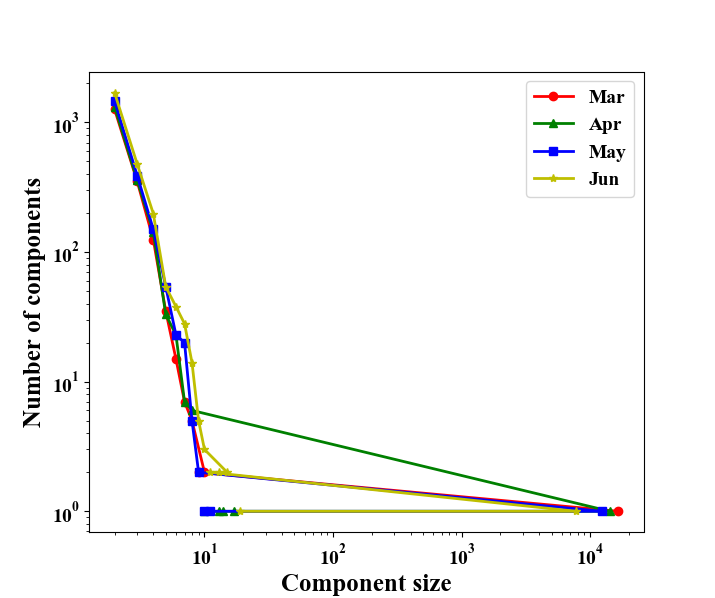}
	\caption{\label{Figure3} (Color online) Component size distribution. The fraction of components with a given component size on a log-log scale. Most nodes are in the largest component.}
\end{figure}

The network characteristics of friendship networks are listed in Table \ref{table2}. From Table \ref{table2}, we uncover that the sizes of friendship networks $S_G$ decline with time, as well as the GCCs $S_{G_1}$, whereas the number of connected components $N$ is increased. These results reflect certain potential characters of friendship formation. That is, to adapt to a new environment, students make as many friends as possible, which leads to the formation of a large friend group. Then, as time goes on and the friendship evolves, the number of friends declines, and more small groups occur, corresponding to the increment of $N$. Moreover, one could observe that the sizes of GCCs $S_{G_1}$ are quite large compared with other components and almost reach the same level of friendship networks $S_G$. For example, on average, the sizes of $S_G$, $S_{G_1}$, and $S_{G_2}$ are $17939$, $12657$, and $14$, respectively. Indeed, this fact can also be observed from Fig. \ref{Figure3}. Unlike small connected components, GCCs are far from them in the double logarithm coordinate of Fig. \ref{Figure3}. Another finding is that the sizes of small connected components follow a power-law distribution $P(x) \sim x^{-\beta}$ with $\beta \approx 3.25$. Due to the above facts, we focus on analyzing the GCCs of friendship networks in this paper if not specified otherwise.  

\begin{table}
	\caption{The size of friendship networks and their components.}
	\label{table2}
	\centering
	\begin{tabular}{cccccc}
		\hline  
		&March&April&May&June&Average\\
		\hline 
		$S_G$&20962&18924&17538&14332&17939\\
		$E_G$&38099&34305&26808&15058&28568\\
		$N$&1229&1897&2105&2523&1939\\
		$S_{G_{1}}$&16489&14180&12226&7734&12657\\
		$S_{G_{2}}$&10&17&11&19&14\\
		$\beta$&3.2&3.3&3.3&3.2&3.25\\
		\hline
	\end{tabular}
\end{table}

\begin{table}
	\caption{The feature values of GCCs of friendship networks from March to June 2019.}
	\label{table3} 
	\centering
	\begin{tabular}{cccccc}
		\hline  
		&March&April&May&June&Average\\
		\hline 
		$\rho \times 10^{-4}$&2.58&3.10&3.12&3.56&3.10\\
		$\left\langle c \right\rangle $&0.16&0.15&0.14&0.13&0.15\\
		$\left\langle L\right\rangle $&6.87&7.01&7.66&9.24&7.70\\
		$\left\langle k \right\rangle $&4.26&4.40&3.81&2.75&3.81\\
		$\alpha$&2.17&2.13&2.29&2.81&2.35\\
		\hline
	\end{tabular}
\end{table}

\begin{figure}[ht]
	\centering
	\includegraphics[scale=0.45]{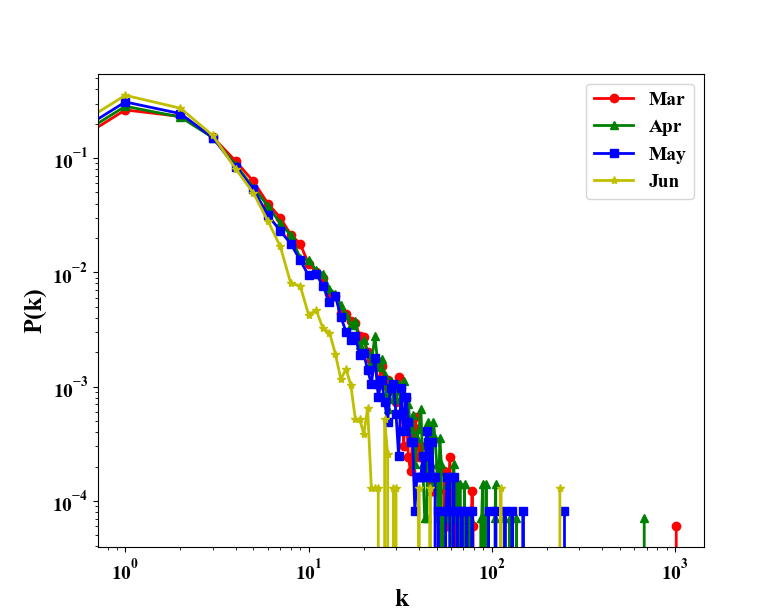}
	\caption{\label{Figure4} (Color online) The degree distribution of friendship networks.}
\end{figure}

Next, the topology parameters of GCCs for various friendship networks are listed in Table \ref{table3}. From the overall point of view, since the network density $\rho$ describes the level of linkages among nodes, the low density value, approximately $3.1 \times 10^{-4}$ on average, illustrates that the friendship network is sparse and the number of friends is limited for most students. The small average shortest path length $\left\langle L \right\rangle $ and large average clustering $\left\langle c \right\rangle $ reveal that the friendship network has a small-world effect. The former indicates that the number of degrees of separation between any two members is small by compared with the size of the population itself. The latter demonstrates that the chance of two students knowing one another is greatly increased, if they have a common acquaintance. In fact, this probability is uniform in a random network, regardless of any two students you choose. The average degree of friendship networks $\left\langle k \right\rangle $ equaling $3.3$ indicates that every student has approximately $3.3$ friends on average. To further study the feature of node degrees, we investigate the characteristic of the degree distribution, and the result is shown in Fig. \ref{Figure4}. One can observe that node degrees follow a power-law distribution $P(r) \sim r^{-\alpha}$ with $\alpha \approx 2.35$. This result indicates that the number of friends is not homogeneous for all the students and that large-degree nodes exist in the friendship network; in other words, some popular students have a large number of friends.  

\begin{figure}[ht]
	\centering
	\includegraphics[scale=0.48]{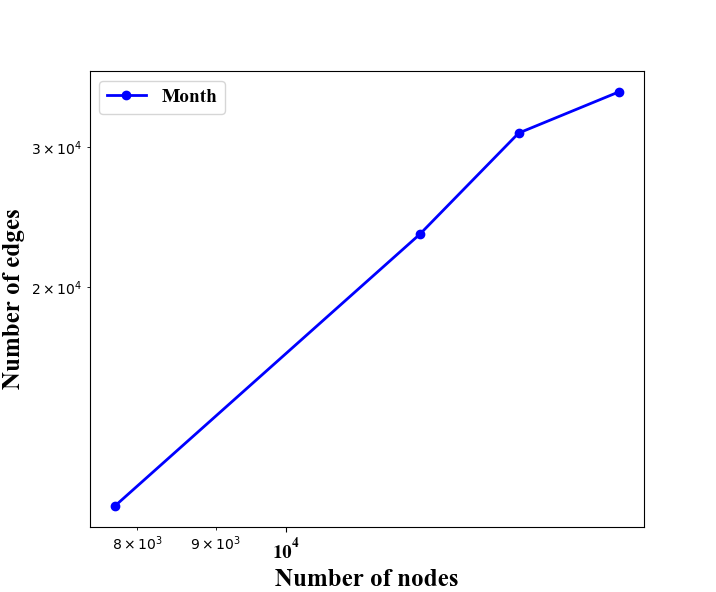}
	\caption{\label{Figure5} (Color online) Number of edges $e$ versus the number of nodes $n$ in log-log scales for several GCCs, which obeys the densification power law with a consistently good fit. Slope: $\gamma = 1.5$.}
\end{figure}

From the perspective of evolution, in Table \ref{table3}, it can be observed that the density $\rho$ and average shortest path length $\left\langle L \right\rangle $ increase with time, whereas the average clustering coefficient $\left\langle c \right\rangle $ and average degree $\left\langle k \right\rangle $ decline. To explain these phenomena, we should know what factors affect these results first. According to the definitions, network density and average degree can be expressed by $\rho=2e/n(n-1)$ and $\left\langle k \right\rangle=2e/n$, respectively. Here, $e$ and $n$ denote the number of edges and nodes of the GCC, respectively. It is necessary to study the relations between nodes and edges. As shown in Fig. \ref{Figure5}, one can discover that they follow the densification power law \citep{Leskovec07graphevolution}:
\begin{equation}\label{eq3}
\begin{split}
e \sim n^{\gamma}, 1<\gamma < 2.
\end{split}
\end{equation}
Particularly, $\gamma = 1.5$, as shown in Fig. \ref{Figure5}. In this case, the network density and average degree are dominated by the number of nodes and edges, respectively. Therefore, when both of them decline, $\rho$ is enhanced and $\left\langle k \right\rangle $ is decreased. Meanwhile, a decrease in the average clustering coefficient $\left\langle c \right\rangle$ indicates that the chance that one's friends are also friends decreases over time. This result can be explained as follows: After the wave of making new friends, only like-minded friends remain. Since the friendship between one's friends is relatively unstable, those edges among neighbor nodes are easier to cut down, which leads to the decline of $\left\langle c \right\rangle$. Furthermore, the average shortest path length $\left\langle L \right\rangle $ will increase since these edges are removed. 

\begin{figure}[ht]
	\centering
	\includegraphics[scale=0.54]{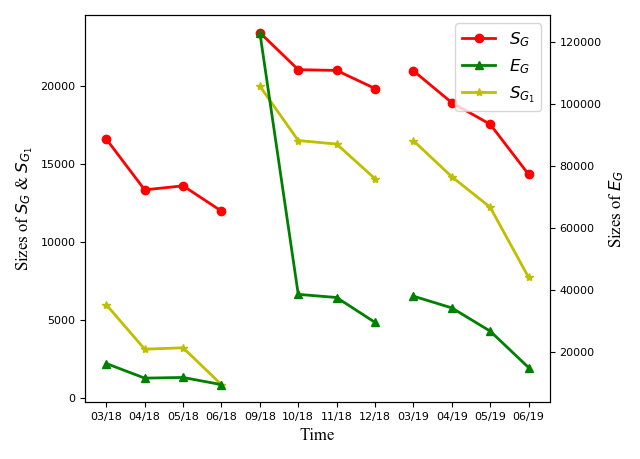}
	\caption{\label{Figure6} (Color online) The evolution behaviors of friendship network in three semesters.}
\end{figure}

Since most universities in China have two semesters in an academic year, where the first is from September to December and the second is from March to June, we enlarged the time frame of data set from March 2018 to June 2019 except for holidays to study the periodicity features. As shown in Fig. \ref{Figure6}, we find that the evolution behaviors of friendship networks are characterized by the semester, that is, networks in different semesters have the similar evolution behaviors. For example, the sizes of friendship networks $S_G$ decline with time in every semester. Another interesting finding is that the size of network edges $E_G$ in September, which was the time new college students enrolled, is much larger than that in other months. According to our observations, the reason of edge increase may come from the contribution of college freshmen, since they have more enthusiasm to make new friends for adapting to a new environment. 

\subsection{Behavior characters}
\subsubsection{Orderliness}
\begin{algorithm}
	\caption{Calculate the actual entropy}
	\begin{algorithmic}[1]
		\Require $data$: Canteen consumption data sequence.
		\Ensure $entropy$: Actual entropy value.\\
		$discrete$: a sequence divided a day into 48 time slices.\\
		$order$: a sequence is converted by a discrete sequence.
		\State Initialize sequence $order$, $entropy$
		\For {$i = data[0]$ to $data[-1]$}
		\State $discrete  \gets data$ 		
		\EndFor
		\State $order \gets num \gets 1$
		\While {$num < len(discrete)$}
		\State Initialize sequence	$m,n,p$
		\State $ 	m,n \gets discrete[:num-1], discrete[num-1:]$
		\For{$i \gets 0$ to $num-1$}
		\If {$m[i] == n[0]$} \State $p \gets i$ \EndIf
		\EndFor
		\If{$len(p) == 0$} \State $order \gets 1$
		\Else
		\For{$j \gets 0$ to $len(p)-1$}
		\State Initialize value	$t \gets 0$
		\State $k \gets p[j]$
		\While{$k < num$}
		\If{$t < len(discrete)-num$ \&\& $m[k] == n[t]$}
		\State  $t \gets t + 1$, 	$k \gets k + 1$
		\Else
		\State break
		\EndIf
		\EndWhile 
		\State $p[j] \gets t$ \EndFor
		\State $q \gets max(p)$
		\If {$q == len(discrete) - num$}
		\State $temp \gets len(discrete) - num + 1$, 	$order \gets temp$
		\Else
		\State $temp \gets q + 1$,	$order \gets temp$
		\EndIf \EndIf
		\State $num \gets num + 1$
		\EndWhile
		\State $entropy \gets $ $ log(len(order))*(order/ sum(discrete))$
		\State \Return $entropy$
	\end{algorithmic}
\end{algorithm}

Student behaviors have very important influences on academic performance, but how to quantify these behaviors is still an open problem in the field of educational research. Orderliness, as a quantitative indicator of describing lifestyle regularity, was first proposed by \cite{Yi2018} via mining large-scale behavioral data, such as taking showers in dormitories and having meals in cafeterias. Here, we use our data set, i.e., individual consumption records in the student canteen, to investigate the features of students' lifestyle regularity and verify the conclusion of orderliness in the study \cite{Yi2018}. 

To this end, we need to obtain the value of the actual entropy first. The calculation method of the actual entropy introduced in Section \ref{orderliness} is presented in Algorithm 2. The time complexity of Calculate the actual entropy is $O(n^2)$, and the space complexity is $O(n^2)$. Because it takes up a lot of RAM, in order to prevent RAM overflow, it is best to reduce the calculation time span. The experimental computer CPU used is Intel i7-9700K, 32GB of RAM, and runs on the Linux distribution Ubuntu18.04 system. The running environment is Pycharm, Anacoda3 and Python3.7. The calculation time of actual entropy on the experimental computer is $47$ minutes.

\begin{figure}[ht]
	\centering
	\includegraphics[scale=0.5]{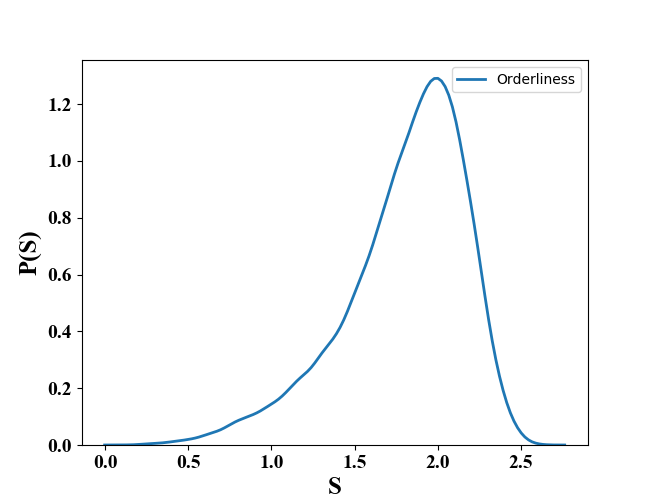}
	\caption{\label{Figure7} (Color online) The distributions of the actual entropy.}
\end{figure}

Using Algorithm 2, we calculate $17,868$ student actual entropy values (considering their GPA can be obtained), and the distribution of the actual entropy is shown in Fig. \ref{Figure7}. One could observe that the actual entropy approximately follows a Gaussian distribution, which is consistent with the result of reference \cite{Yi2018}. For most students, the value of the actual entropy is approximately $2.0$, and only a few students have small values. According to the relationship between orderliness and actual entropy, the smaller the actual entropy is, the stronger the orderliness is. Our results demonstrate that high-orderliness students are limited and that most students are distributed at mid-level positions.

\begin{figure}[ht]
	\centering
	\includegraphics[scale=0.5]{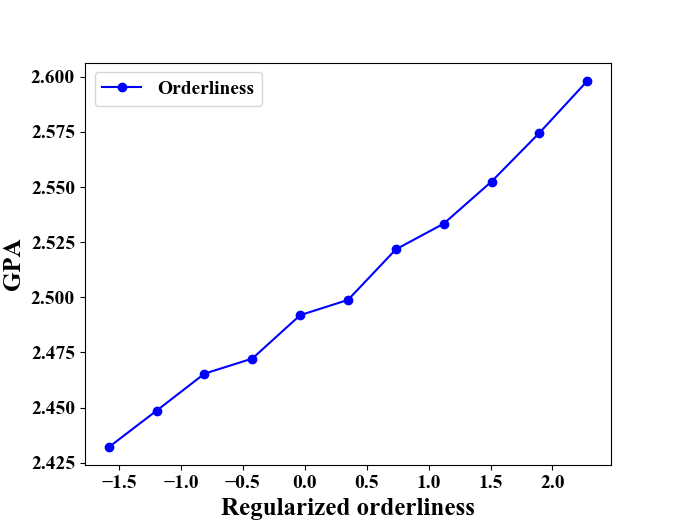}
	\caption{\label{Figure8} (Color online) Relationship between orderliness and GPA. Binned statistics are used to aggregate the data points, where regularized orderliness is divided into $11$ bins. The mean value of data points in each bin is presented. Spearman's rank correlation coefficients for GPA-Orderliness ($r=0.155$, $p<0.0001$) suggest the statistical significance.}
\end{figure}

To investigate the relationship between orderliness and academic performance, we first calculate all the student GPAs, and then divide the students into $11$ different groups according to their value of orderliness. The gap of each group is the same. The average value of the GPA in each group is calculated, and the results are presented in Fig. \ref{Figure8}. It can be easily observed that orderliness is positively correlated with GPA, implying that high-orderliness students might achieve better academic performance, which again demonstrates the results of the study of reference \cite{Yi2018}. 

Furthermore, to detect the statistical significance, we choose the same number of students from each group and apply Spearman's rank correlation coefficient to quantify the correlation coefficient between the GPA and orderliness. In fact, the value of Spearman's rank correlation coefficient is in the range $\left[ 0,1\right] $, and the large absolute value reflects a high correlation. Our result of $r=0.155$ indicates that there is significant correlation between the orderliness and GPA.

\subsubsection{Diligence}
Another parameter describing student behavior is diligence, which is measured by the frequency of entering the library. To investigate the features of student diligence, we collected approximately $5.6$ million records of these behaviors in the library, and obtained the values of diligence for $26,753$ students after matching with GPAs.

\begin{figure}[ht]
	\centering
	\includegraphics[scale=0.45]{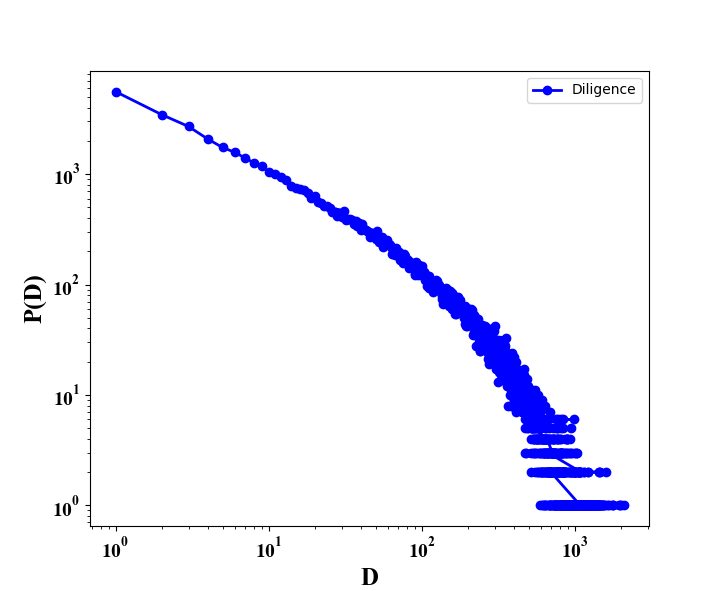}
	\caption{\label{Figure9} (Color online) The distribution of the number of times entering/existing the library. }
\end{figure}

First, we investigate the feature of the diligence frequency distribution, and the result is shown in Fig. \ref{Figure9}. Unlike orderliness, the diligence follows a power-law distribution, which means that the diligence is heterogeneous for different students. That is, even though most students seldom go to the library, a few still often go there and spend much time learning. This phenomenon can be explained by the fact that, for most students, the purpose of going to library might be to deal with examinations or to conduct self-learning prior to the exam. Nevertheless, for some other students, it is a habit, and there are many records of when they enter the library. Another explanation is that human behaviors have a memory effect; the more times you go to the library, the more likely you would do it next time.

\begin{figure}[ht]
	\centering
	\includegraphics[scale=0.5]{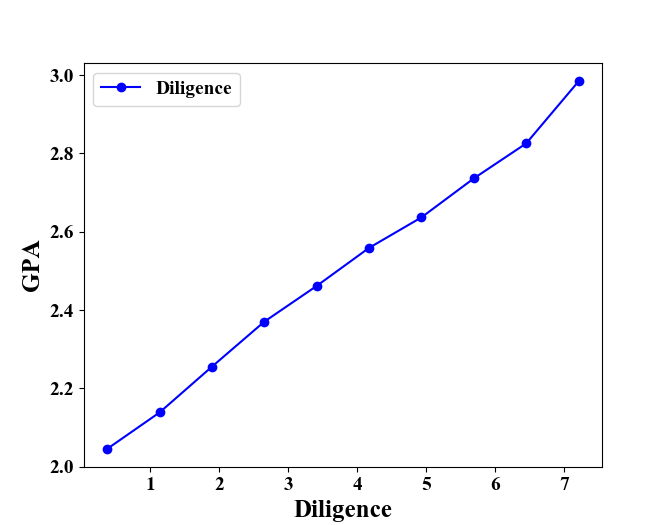}
	\caption{\label{Figure10} (Color online) Relationship between the diligence and GPA. Log-binned statistics are used to aggregate the data points. The mean value of data points in each bin is presented. Spearman's rank correlation coefficients for GPA-Diligence ($r=0.374$, $p<0.0001$) suggest statistical significance.}
\end{figure}

Intuitively, the more diligent a student is, the better his/her academic performance is. To prove this conclusion, we collect student GPAs and analyze the relationship between the diligence and GPA.  The result is shown in Fig. \ref{Figure10}. From Fig. \ref{Figure10}, one can observe that the diligence is positively correlated to GPA, which confirms our statement. Furthermore, we obtain a high level of statistical significance for this relationship, and the value of Spearman's rank correlation coefficient $r$ equals $0.374$, which reflects a high correlation. 

\subsubsection{Assortativity analysis}
The behavior characters such as orderliness and diligence have been studied in the above sections, but they are just viewed from an individual perspective. In fact, student behaviors are always influenced by others; for example, a student will go to library if his/her friend invited him/her. To investigate the relations between behavior characters and student interactions, based on friendship networks, we use equation (\ref{eq4}) in section \ref{assortativity} to calculate the assortativity coefficients of the node degree, orderliness, diligence and GPA. All the results are listed in Table \ref{table4}.

First, the result of the assortativity coefficient of node degree $r_{degree}$ indicates that friendship networks have no degree correlations since $r_{degree}$ nearly equals zero, which implies that whether or not any two students are friends is independent of the number of their friends. 

Unlike the result of the node degree, the values of behavioral assortativity coefficients, orderliness $r_{orderliness}$ and diligence $r_{diligence}$, are larger than zero, especially the orderliness, whose value almost equals $0.4$. These results indicate that friendship networks are assortative for orderliness and diligence, which are also called peer effects. That is, a regular student tends to make friends with regular behaviors, and friends of a diligent student are also diligent. Furthermore, the assortativity coefficient of the GPA is calculated and its result is almost the same as the orderliness, which indicates that friendship networks are also assortative for student GPAs. In conclusion, friendship networks have strong peer effects on individual behaviors even though networks have no degree correlations.

\begin{table}
	\caption{ Values of assortativity coefficient of the node degree, orderliness, diligence and GPA.}
	\label{table4}
	\centering
	\begin{tabular}{cccccc}
		\hline  
		&March&April&May&June&Average\\
		\hline 
		$r_{degree}$&-0.02&-0.02&0.03&-0.02&-0.01\\
		$r_{orderliness}$&0.37&0.36&0.38&0.46&0.39\\
		$r_{diligence}$&0.27&0.24&0.25&0.27&0.26\\
		$r_{GPA}$&0.41&0.40&0.42&0.46&0.42\\
		\hline
	\end{tabular}
\end{table}

\subsubsection{Percolation analysis}
With regard to mesoscale properties, assortativity analysis in friendship networks has been shown to have strong peer effects on individual behaviors. In addition to mesoscale features, it is also necessary to investigate the effects of global network topology on student behaviors. To this end, we adopt percolation theory to analyze the characteristics of orderliness on friendship networks. 

First, we record the value of orderliness on every node. Then a threshold $m$, which varies from the minimum to the maximum of orderliness, is defined to act as the only control parameter in our percolation analysis. The $s_i$ of the node with orderliness $o_i$ will be grouped into one of two classes: orderly state for $o_i \ge m$ or disorderly state for $o_i < m$, i.e.\\
\begin{equation}
s_{i} =
\begin{cases}
1 & \text{$o_{i} \ge m$}\\
0 & \text{$o_{i} < m$}
\end{cases}
\end{equation} 

\begin{figure}[ht]
	\centering
	\includegraphics[scale=0.34]{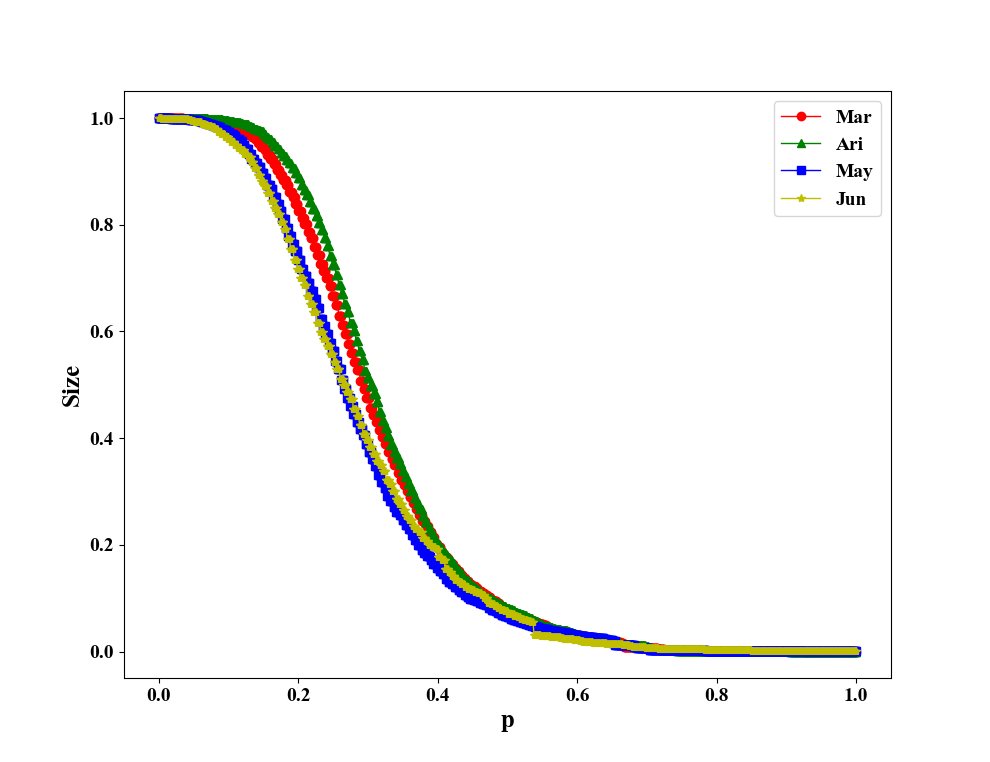}
	\caption{\label{Figure11} (Color online) The percolation of orderliness on various friendship networks.}
\end{figure}

\begin{figure}[ht]
	\centering
	\includegraphics[scale=0.25]{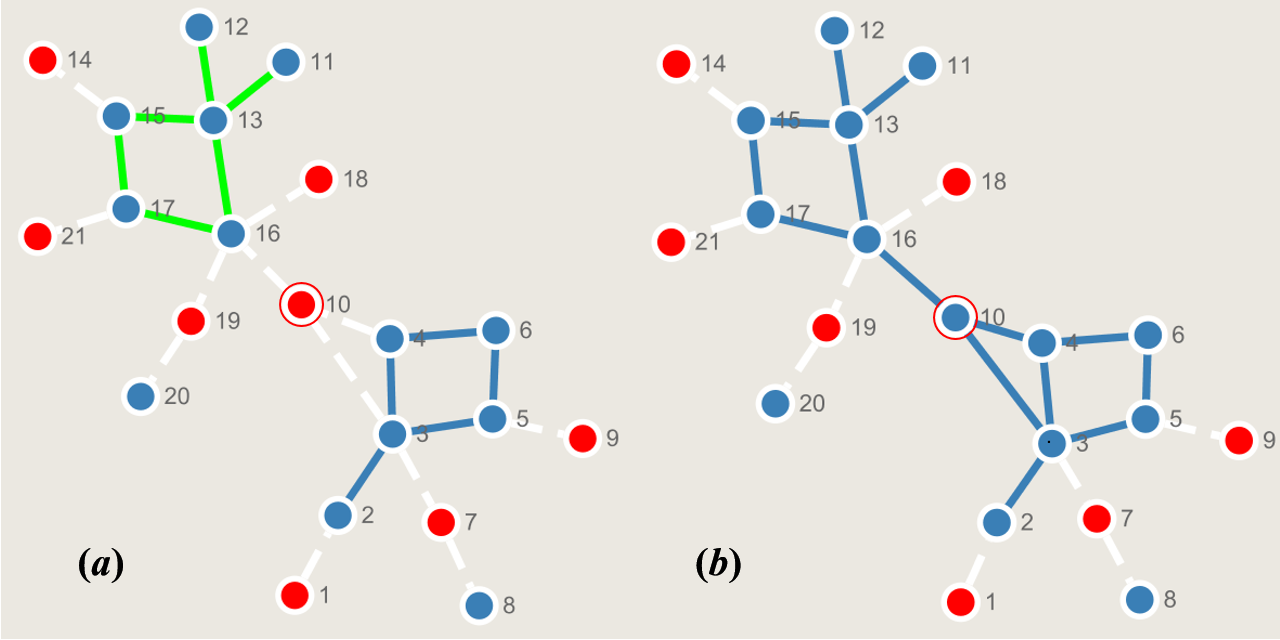}
	\caption{\label{Figure12} (Color online) Schematic of the key nodes in friendship networks. The blue/red nodes are orderly/disorderly states, and the solid/dotted lines denote nodes connected/disconnected by edges. (A) A typical example of a friendship network just above criticality, where some nodes are disorderly states at criticality. Removal of them will disintegrate the giant connected component as two components which connected by blue edges and green edges, respectively. (B) The same friendship network after addition of the key node $10$ (circled in red), where the giant connected component occurs again.}
\end{figure}

The fraction of orderly state nodes ($s_i = 1$) in the network, which can be considered the occupation fraction $p$ in percolation, increases as we gradually reduce the threshold $m$. In this process, we choose an appropriate interval ($\Delta m = 0.01$ in our study) so that the decrease in $m$ is small enough to keep at most one node's state modified ( one $s_i$ changes from $0$ to $1$). As we decrease the threshold $m$, components of orderly state nodes ($s_i = 1$) will emerge, and we can observe the occurrence of the percolation process in the network. For large $m$, almost all the nodes are considered as disorderly and only small components will appear. For small $m$, small components will merge into larger components, showing the organization process of local orderliness. The emergence of the GCC indicates the occurrence of the percolation transition and the formation of global orderliness. At a certain fraction of orderly state nodes $p_c$ (determined by $m_c$), the second largest component reaches its maximum, which signifies the dissipation of global orderliness. According to percolation theory, this value $p_c$ refers to the critical threshold of global orderliness percolation. 

Fig. \ref{Figure11} shows the percolation process of orderliness on the GCCs of friendship networks. As shown in Fig. \ref{Figure11}, one can observe that the critical threshold $p_c$ is approximately equal to $0.2$, below which almost all the students are in an orderly state. However, when $p$ is larger than $p_c$, the GCC composed of orderly state nodes is divided into many small components, and the size of the largest component is far less than that of the network. These results give us an important inspiration, that is, to achieve swarm intelligence, and the best way is to develop/intervene the key nodes near the critical threshold. For instance, the left-hand chart in Fig. \ref{Figure12} is a typical example of a friendship network just above criticality, where some nodes are disorderly states at criticality. In this case, the GCC is disintegrated into two small connected nodes by a key node $10$ circled in red in Fig. \ref{Figure12}. Compared to changing other nodal states, the best way of reoccurring the GCC is to develop the orderliness of node $10$ (see right-hand chart). Therefore, those key nodes identified at $p_c$ can provide opportunities to significantly improve the global network orderliness with a minor cost.

\section{Conclusions}
In this paper, we studied the evolution features of friendship and relations between behavior characters and student interactions from the perspective of network topology by mining behavioral data on campus. To infer friend ties, we proposed a theoretical framework to determine the critical value of the co-occurrence frequency and found that the critical value for a month should be five since any pair of strangers cannot be found in terms of statistical expectations. The self-report results further confirmed the validation of our method of inferring friend ties. Moreover, we investigated the functional relationship between the network size and critical value, and there was a significant exponential relationship between them. This finding indicated that a reasonable critical value was of importance because a larger value would omit many real friends or a smaller value would mistake many strangers for friends.

To investigate the evolutionary nature of friendship, the topological characteristics of four friendship networks and their GCCs were analyzed. First, we discovered that friendship networks were sparse networks, node degrees followed a power-law distribution, and the small-world effect existed due to the high clustering and short path lengths. Second, we found that the sizes of both friendship networks and GCCs declined with time, whereas the number of small connected components increased, which described the phenomenon that a large friend group decayed into many small groups. Next, the network density and average shortest path length increased with time, whereas the average clustering coefficient and average degree declined. According to the results of topology parameters, we may safely arrive at the conclusion that students make many friends when they are in a new environment; as time goes on, the number of friends will decrease, only like-minded friends will remain, and the friendship among one's friends is not stable. Moreover, we found that the evolution behaviors of friendship networks are characterized by the semester, and the size of network edge in September is much larger than that in other months due to the enrollment of new students. 

We then studied the relations between behavior characters and student interactions. First, we investigated the characteristics of orderliness and diligence. The results indicated that both orderliness and diligence were positively correlated with the GPA; however, the former followed a Gaussian-like distribution and the latter followed a power-law distribution. Next, we calculated the assortativity coefficient of the degree, orderliness, diligence and GPA in friendship networks, and the results indicated that friendship networks had strong peer effects on student behaviors. Using percolation theory, we investigated the effects of global network topology on behavior characters, and the results showed that a phase transition of orderliness existed in friendship networks. Since orderliness is strongly related to student academic performance, this result might provide some reference for managers to intervene those crucial students and realize the goal of swarm intelligence.

According to above results, some advises could be provided to help educators in their study. On the one hand, it is best to schedule most social and academic activities for each student timely in a semester, so that students have the best motivations to make new friends during this time. On the other hand, we could provide some methods to educational administrators for early warning of at-risk students with potential low academic performances. For example, educators could set the critical threshold of the orderliness and diligence in the student card system, and students will be kindly reminded if their values were reduced below the critical threshold.

\section{Limitations and further research}
This study has several limitations, which can provide a direction for future research. Firstly, although the research data in this study came from different behaviors on campus in China, the study's findings may be restricted to the Chinese students. Further research should extend to other countries' university and compare the effects of friendship network characteristics on behaviors. Secondly, from the perspective of network evolution, the size of time window may influence friendship network formation, and adopting the sliding-time-window technology would observe more detailed evolution features of friendship formed. Thus, future research should vary or slide time window and study the microscopic mechanism of friendship formed. Third, some personality traits were not considered in this study. In Section \ref{method of constructing network}, we hypothesized that everyone chooses the canteen window randomly, nevertheless, the personal consumption habits were not considered. Therefore, it is necessary to analyze the effects of  personality traits on friendship networks.

In our study we have got the conclusion that high-orderliness students would achieve better academic performance, but some studies also found that people who have a little more disorder would be more creative and resilient \citep{Tim2017,KIM201715}. Another interesting research demonstrates that small teams produce significantly more disruptive work than large teams \citep{Wu2019}. Based on these facts, the directions of further research can be extended to investigate the creativity of low-orderliness students and small connected components of friendship networks, which accounts for around $30\%$ but we have neglected in this paper. 

\section{Declaration of Competing Interests}
The authors declare that they have no known competing financial interests or personal relationships that could have appeared to influence the work reported in this paper.

\section{Credit authorship contribution statement}
\textbf{Z. Yang}: Supervision, Project administration. \textbf{Z. Su}: Data curation, Formal analysis, Investigation, Writing - original draft, Writing review \& editing. \textbf{S. Liu}: Conceptualization, Supervision, Writing - review. \textbf{Z. Liu}: Conceptualization, Writing - review \& editing. \textbf{W. Ke}: Conceptualization, Methodology, Investigation, Writing - original draft. \textbf{L. Zhao}: Data collection, Writing - review \& editing.

\section{Acknowledgments}
This work was supported by the National Natural Science Foundation of China (Grant Nos. 61937001, 61977030) .\\
\bibliographystyle{model5-names}
\biboptions{authoryear}
\bibliography{ref-students}
\end{document}